# Direct measurement of terahertz conductivity in a gated monolayer semiconductor

Su-Di Chen[1,2,3], Qixin Feng[1,3], Wenyu Zhao[1], Ruishi Qi[1], Zuocheng Zhang[1], Dishan Abeysinghe[1,3], Can Uzundal[1,3,4], Jingxu Xie[3,5], Takashi Taniguchi[6], Kenji Watanabe[7], Feng Wang[1,2,3,*]

[1] Department of Physics, University of California, Berkeley, CA 94720, USA.

[2] Kavli Energy NanoSciences Institute, University of California Berkeley and Lawrence Berkeley National Laboratory, Berkeley, CA 94720, USA.

[3] Materials Sciences Division, Lawrence Berkeley National Laboratory, Berkeley, CA 94720, USA.

[4] Department of Chemistry, University of California, Berkeley, CA 94720, USA.

[5] Graduate Group in Applied Science and Technology, University of California, Berkeley, CA 94720, USA.

[6] Research Center for Materials Nanoarchitectonics, National Institute for Materials Science, 1-1 Namiki, Tsukuba 305-0044, Japan.

[7] Research Center for Electronic and Optical Materials, National Institute for Materials Science, 1-1 Namiki, Tsukuba 305-0044, Japan.

[*] To whom correspondence should be addressed: fengwang76@berkeley.edu

**Abstract:** Two-dimensional semiconductors and their moiré superlattices have emerged as important platforms for investigating correlated electrons. However, many key properties of these systems, such as the frequency-dependent conductivity, remain experimentally inaccessible because of the mesoscopic sample size. Here we report a technique to directly measure the complex conductivity of electrostatically gated two-dimensional semiconductors in the terahertz frequency range. Applying this technique to a WSe$_2$ monolayer encapsulated in hBN, we observe clear Drude-like response between 0.1 and 1 THz, in a density range challenging to access even in DC transport. Our work opens a new avenue for studying tunable van der Waals heterostructures using terahertz spectroscopy.

**Main text:** Conductivity as a function of frequency is a key observable in solids, encoding rich information about the underlying electronic states [1]. In correlated electron systems, conductivity in the terahertz (THz) frequency range is of particular interest because of its overlap with emergent energy scales from quasiparticle scattering rates in metals to energy gaps and collective modes in broken-symmetry states [2–5]. Experimentally, THz spectroscopy, a direct probe of THz conductivity, has played vital roles in the study of correlated states hosted by bulk materials and large thin films. However, its application to van der Waals heterostructures, a rising platform [6–10] featuring exceptional tunability from mechanical stacking and electrostatic gating, has been challenging. First, the size of the vdW heterostructures, typically around tens of microns, is much smaller than the wavelength of the THz photons on the order of hundreds of microns. This length-scale mismatch necessitates near-field approaches [11–14], through which to extract quantitatively the conductivity spectra is often difficult. Moreover, as the sample and the gate electrode are arranged in a parallel-plate-capacitor geometry, it is hard to separate the THz response of the sample from that of the gate material.

On-chip THz spectroscopy is an emerging technique that partially resolves these challenges [15–19]. In this technique, the sample interacts with THz fields confined in waveguides. If the sample is conducting enough, simple circuit models can be used to convert the measured waveguide transmission to sample conductivity [15]. Background contributions from the gate electrode can also be ignored, if its conductance is much lower than that of the sample. Because of these constraints, absolute measurement of THz conductivity in a vdW heterostructure has only been accomplished in graphene at high dopings, using a much-less-conducting transition metal dichalcogenide (TMD) as gate [15]. In this work, we will demonstrate THz conductivity measurement on gated monolayer TMD for the first time. By introducing an improved waveguide design, a new gate material, and a new analysis protocol, our work opens up possibilities for the THz study of generic vdW heterostructures.

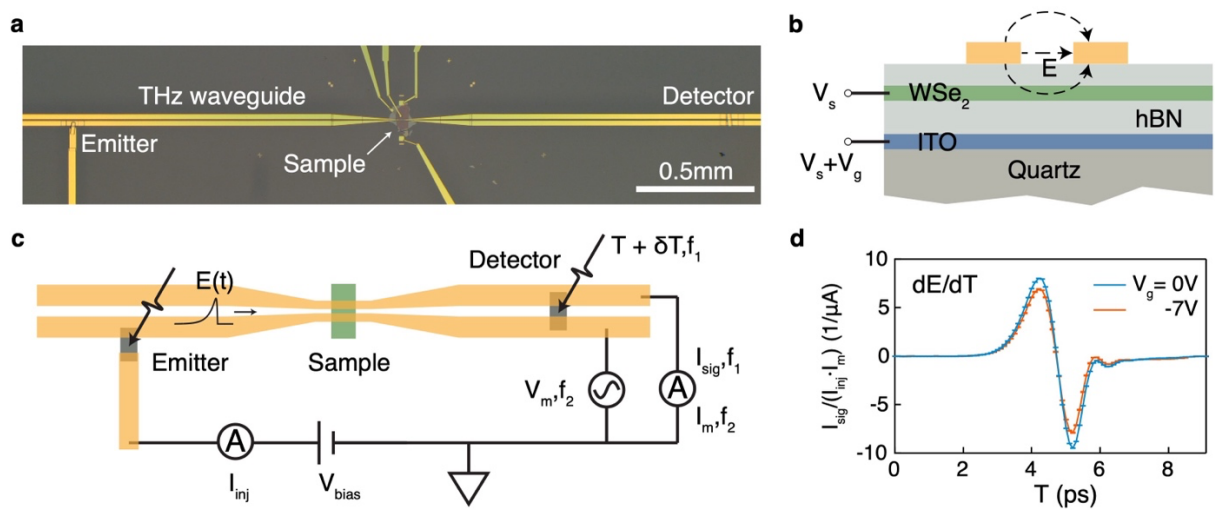

Fig. 1. Terahertz (THz) time-domain spectroscopy inside an on-chip waveguide. **a**. Microscope photograph of the center part of the device for THz spectroscopy. **b**. Schematic cross-sectional view of the sample segment in **a** perpendicular to the THz propagation direction. The monolayer $WSe_2$ sample is gated by a 5-nm-thick indium tin oxide (ITO) electrode across a 73-nm-thick hexagonal boron nitride (hBN) dielectric layer. **c**. Schematics of the measurement circuit. Pulsed THz fields (E) with few-ps duration are generated at 80 MHz repetition rate by a laser-excited photoconductive switch (emitter), and after interacting with the sample, sampled at different time delays (T) by a second switch (detector). T is modulated by $\delta T = 0.17$ ps at $f_1 = 50$ Hz, and a small AC monitoring voltage $V_m$ is applied between the waveguide traces at $f_2 = 127$ Hz. In-phase photocurrents across the detector at $f_1$ (denoted $I_{sig}$) and $f_2$ (denoted $I_m$), and the average photocurrent injected through the emitter ($I_{inj}$) are measured. **d**. Normalized signal $I_{sig}/(I_{inj} \cdot I_m)$, which is proportional to dE/dT of the transmitted pulse convolved with the detector time resolution, as a function of T for undoped (gate voltage $V_g = 0$ V) and hole-doped $WSe_2$ ($V_g = -7$ V). Error bars represent 1-$\sigma$ noise in measurement.

The layout of our device is shown in Figs.1a and b (see also Fig. S1). The topmost layer of the structure is a gold coplanar stripline (CPS) waveguide [20] with integrated photoconductive switches made of low-temperature-grown GaAs [2]. Broadband THz pulses are injected into the CPS by exciting the biased emitter switch with a femtosecond laser pulse. The transmitted THz electric field is then sampled in the time domain by the detector switch triggered with a second laser pulse. The total length of the CPS is 8.5 mm, with the emitter and detector placed at its trisecting points. This arrangement ensures that the directly transmitted signal, which we collect and analyze, is separated in time from those reflected by the switches and the CPS terminations.

Near the center of the CPS, the width of and gap between the metal traces are gradually narrowed to 2.26 and 0.49 μm, respectively. The sample, a monolayer $WSe_2$ flake encapsulated in hexagonal boron nitride (hBN), sits right beneath the narrowest segment. This tapered design enhances the THz field in the sample, and thus the sensitivity of the measurement. Moreover, it allows the contacts and the edge of the flake to be placed at distances much larger than the CPS dimension perpendicular to the propagation direction, effectively eliminating their interference with the waveguide mode.

The sample is charge neutral and insulating at low temperature. We define an active region smaller than the $WSe_2$ flake using a rectangular-shaped backgate 20-um wide along the propagation direction (see supporting information section 1 for details). The gate electrode is made of indium tin oxide (ITO), an electron-doped semiconductor, with a thickness around 5 nm. Through DC transport measurements, we determine its sheet conductance ($\sigma_{ITO}$) and mobility ($\mu_{ITO}$) to be around 0.11 mS and 40 $cm^2V^{-1}s^{-1}$, respectively (see supporting information section 4 for details). These properties are ideal for our measurements. First, assuming an electron effective mass of $0.3m_0$ [21], $\mu_{ITO}$ translates to a transport scattering time $\tau_{ITO} \approx 7$ fs. Thus, below a few THz, $\sigma_{ITO}$ can essentially be treated as frequency independent, simplifying the modeling of the structure. Moreover, $\sigma_{ITO}$ is small enough, such that the gate electrode does not strongly reflect or screen the THz field which needs to interact with the sample (see supporting information section 8 for more discussions). In addition, the electron density in ITO is estimated to be $n_{ITO} \approx \sigma_{ITO}/e\mu_{ITO} \approx 1.7 \times 10^{13}$ $cm^{-2}$, which sets a generous upper bound for the maximal achievable electron density in the sample through gating (the hole density is not limited). Here e denotes the elementary charge.

We illustrate our measurement circuits in Fig.1c. A femtosecond laser is used to excite first the emitter and after a time delay T the detector, at a repetition rate of 80 MHz. The emitter is biased at $V_{bias}$ = 7.6 V using a source meter, which also measures the average photocurrent injected into the CPS, $I_{inj}$ (around 80 nA). The photocurrent through the detector, which is proportional to the transient transmitted electric field E(T) cross-correlated with the time resolution of detection, is collected by a current preamplifier. To reduce noise, we modulate T with an amplitude of 0.17 ps at $f_1$ = 50 Hz, and read out the resulting modulation in photocurrent $I_{sig}$ using a lock-in amplifier, effectively measuring dE(T)/dT. In addition, we apply a small AC voltage, $V_m$ = 2 mV at $f_2$ = 127 Hz, between the waveguide traces and monitor the resulting in-phase detector photocurrent $I_m$ using a second lock-in amplifier. As the photoconductances of the emitter and detector are sensitive to drifts in laser power and beam positions, we normalize $I_{sig}$ by $I_{inj} \cdot I_m$ to remove extrinsic drifts from the signal.

During measurements, we change the gate voltage $V_g$ between the sample and the ITO to access different doping states. We also set the voltage difference between the sample and the waveguide to $V_s = -0.2$ V, to align the hole density underneath the metal traces with the rest of the gated region (see supporting information section 4 for details). Two representative time domain traces of $I_{sig}/(I_{inj} \cdot I_m)$ measured at 20 K are shown in Fig. 1d, at $V_g = 0$ V ($WSe_2$ undoped) and -7 V ($WSe_2$ hole-doped). Because $I_{sig}/(I_{inj} \cdot I_m) = C_0 \cdot dE/dT \star R$, where $C_0$ is a constant, $\star$ denotes cross-correlation, and R represents the finite resolution of detection, after Fourier transform (FT) we obtain $t/t_0 = FT[I_{sig}/(I_{inj} \cdot I_m)]/ FT[I_{sig}/(I_{inj} \cdot I_m)]_{V_g = 0V}$ in the frequency domain, where t ($t_0$) is the field transmission coefficient in the doped (undoped) state.

We plot the amplitude and phase of $t/t_0$ as functions of frequency in Figs. 2f and g, respectively. The remaining task is to convert $t/t_0$ to the conductivity $\sigma = \sigma_1 + i\sigma_2$ of doped holes. Instead of using approximate analytical formulas, we numerically obtain the mapping between $t/t_0$ and $\sigma$. First, we discretize the CPS into a large number of segments (Fig. 2a) where each segment has its own cross-sectional geometry. We have characterized the actual geometries using optical microscopy, scanning electron microscopy (SEM), and atomic force microscopy (AFM) to ensure the accuracy of modeling. For each segment at every frequency, we solve the two-dimensional waveguide mode numerically (see Figs. 2b and c for example, see also supporting information section 7 for more details) and obtain its effective refractive index $n_{eff}$ and characteristic impedance $Z_0$. In particular for the sample segment, we solve the mode on a parameter grid spanned by different $\sigma_1$ and $\sigma_2$ values while setting $\sigma_{ITO}$ to its experimentally measured value at $V_g = 0$ V. With all $n_{eff}$ and $Z_0$ known, we then use a transfer matrix calculation [22] to determine the waveguide transmission t as a function of $\sigma_1$ and $\sigma_2$. This is analogous to calculating the optical transmission through a multilayer medium, although here in the CPS each segment is parameterized by both $n_{eff}$ and $Z_0$, instead of a single refractive index.

Using these calculations, we can obtain $t/t_0 = t(\sigma)/t(\sigma=0)$ on a grid in the complex plane of $\sigma$ at every frequency (see Figs. 2d and e for an example at 0.5 THz). Within the relevant $\sigma$ range, both the amplitude and phase of t/t$_0$ evolve in a nearly monotonic fashion, with their constant-value contour lines following distinct directions. This indicates a one-to-one correspondence between t/t$_0$ and σ. We can thus convert the experimentally measured data to conductivity using the simulated mapping. As a remark, we have ignored the change of $\sigma_{ITO}$ with V$_g$ in this analysis because $\mu_{ITO}$ is much small than the mobility of WSe$_2$ within the measured frequency range. If needed, one can also take this effect into account by calculating the conversion mapping using the measured $\sigma_{ITO}$ values at each V$_g$ instead (see supporting information section 5).

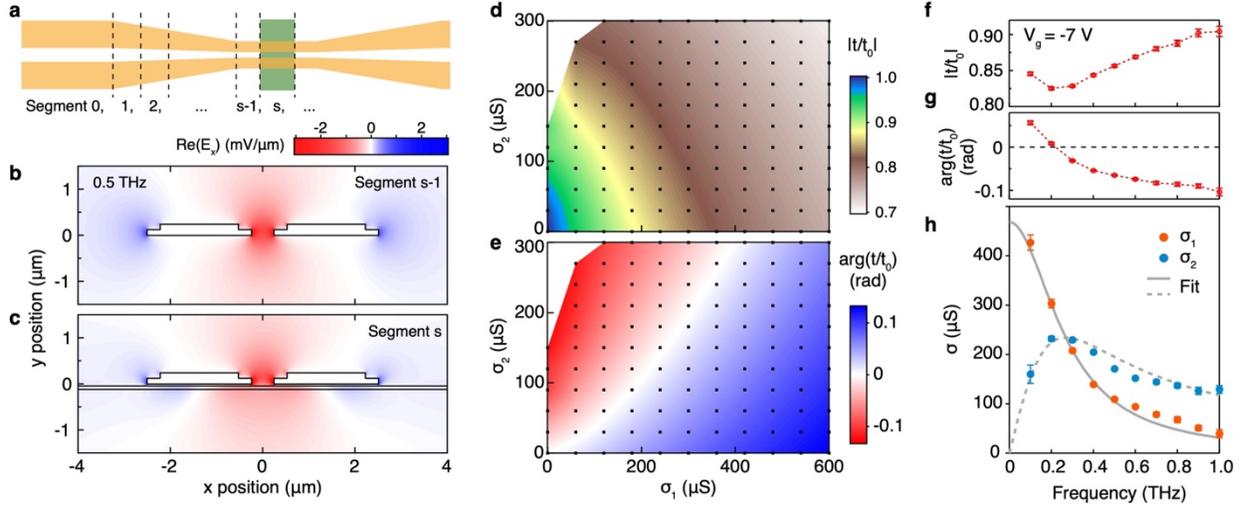

Fig. 2. Extract conductivity from experiment. **a**. Schematic of waveguide modeling. The structure is discretized into many segments. The waveguide mode is numerically solved for each segment, allowing the total transmission to be calculated using transfer matrix method. The actual segment interval is 0.5 μm in the tapered section, which is exaggerated in the schematic for clarity. **b**. Cross-sectional mode profile at 0.5 THz in segment s-1, near the metal traces outlined in black. The color scale represents the in-phase component of electric field along the horizontal direction and is oversaturated for clarity. Data are normalized such that the voltage difference between the metal traces equal to 1 mV. **c**. Same as **b** for segment s. The horizontal lines mark the WSe$_2$ (top) and ITO (bottom) layers, with their sheet conductance at 300+300i μS and 109 μS, respectively. **d**. Amplitude of the transmission coefficient t at 0.5 THz as a function of the real ($\sigma_1$) and imaginary ($\sigma_2$) parts of sample conductance normalized by that at $\sigma_1=\sigma_2=0$ ($t_0$). Actual simulations are performed on the grid marked by black dots and the color plot is a Voronoi interpolation of the simulations. **e**. same as **d** for the phase shift arg(t/t$_0$). **f**. Frequency evolution of |t/t$_0$| obtained from measured data in Fig. 1**d**, using the undoped trace as reference. **g**. same as **f** for arg(t/t$_0$). **h**. Sample conductance extracted from data in **f** and **g**, using simulated mappings at 10 frequencies. A simultaneous Drude fit to both the real and imagery parts are plotted in grey. Error bars in **f**, **g**, and **h** represent 1-σ error from noise in measurements.

The converted results from V$_g$ =-7 V are plotted in Fig. 2h. We observe a clear drop of $\sigma_1$ with increasing frequency, and a non-monotonic $\sigma_2$ that crosses $\sigma_1$ near 0.3 THz. We fit both curves simultaneously using a Drude model with a frequency-independent scattering time [1]: $\sigma = \frac{D}{\pi} \frac{1}{1/\tau - i \cdot 2\pi f}$. Here the only fitting parameters are D and τ, which denote the Drude weight and the scattering time, respectively. We obtain D = 2.48(6)×10$^9$ S/s and τ = 0.59(2) ps. The fitting results are plotted in grey, which well reproduce the data.

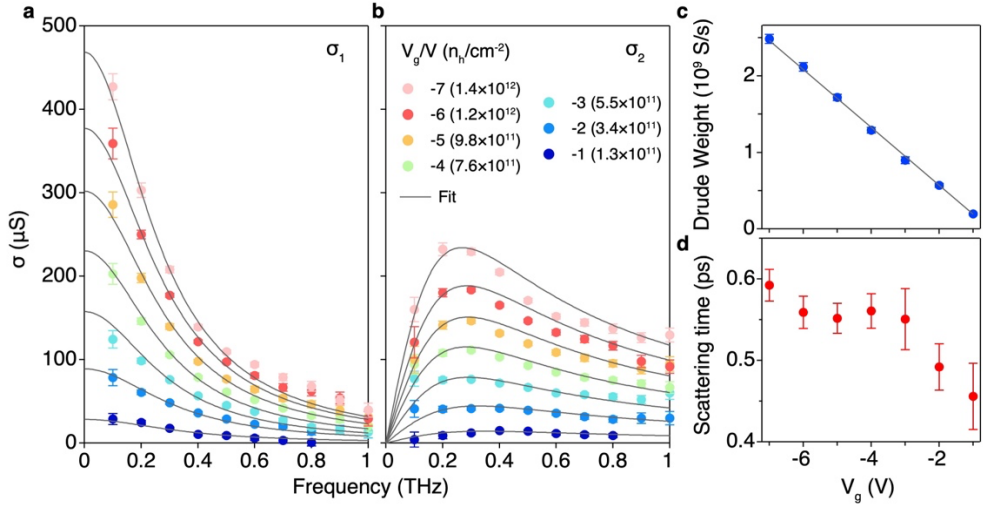

Fig. 3. Doping evolution of conductivity spectra. **a**, **b**, real and imaginary parts of WSe$_2$ conductance as functions of frequency at various $V_g$ (labeled in **b** in units of V). The corresponding hole densities $n_h$ are calculated using a parallel-plate-capacitor model and provided in parentheses in units of cm$^{-2}$. For each $V_g$, data from both panels are fit to the Drude expression and the results are plotted in grey. Error bars in **a** and **b** represent 1-σ error from noise in measurements. **c**. Drude weight extracted from the fitting in **a** and **b** as a function of $V_g$. A linear fit to the data is plotted in grey. **d**. Same as **c** for the scattering time. Error bars in **c** and **d** represent 1-σ error from fitting.

We show the doping evolution of σ in Fig. 3. All data are measured at 20 K, the base temperature of our cryostat. We determine the hole density as $n_h = \varepsilon_0\varepsilon_r(V_g-V_0)/de$, where $\varepsilon_0$ is the vacuum permittivity, $\varepsilon_r$ = 2.8 is the DC dielectric constant of hBN along the out-of-plane direction, $V_0$ = -0.4V is the onset gate voltage of hole doping, and d = 73 nm is the thickness of the hBN dielectric measured by AFM. Here $V_0$ is determined using optical and ITO sensing methods on the same device, and $\varepsilon_r$ is measured on another hBN flake exfoliated from the same batch of hBN crystals (see supporting information sections 3, 4, and 6 for details). We observe that the conductivity below 1 THz in the density range between 1.3×10$^{11}$ and 1.4×10$^{12}$ cm$^{-2}$ well follows a canonical Drude form. The Drude weight is found to evolve linearly as a function of $V_g$ (Fig. 3c), which indicates a constant effective mass in the measured doping range. A linear fit to the data extrapolates to 0 at $V_g$ = -0.5 V, consistent with $V_0$ within experimental uncertainties. From the slope of the fit, we obtain a hole effective mass m* = πe$^2$(dD/d$n_h$)$^{-1}$ = 0.50 ± 0.09 m$_0$, where m$_0$ is the bare electron mass and the uncertainty mostly originates from that in $n_h$. This is close to values reported in previous quantum oscillation measurements [23,24], and agrees with band structures measured by angle-resolved photoemission spectroscopy without magnetic fields [25].

Fig. 3d shows the extracted scattering time, which gradually reduces from 0.59(2) ps at $V_g$ = -7 V to 0.46(4) ps at -1 V. These values correspond to mobilities of 2.1×10$^3$ and 1.6×10$^3$ cm$^2$V$^{-1}$s$^{-1}$ at $n_h$ = 1.4×10$^{12}$ and 1.3×10$^{11}$ cm$^{-2}$, respectively. Remarkably, DC transport measurements in this density range have only become possible very recently, after the development of low-resistance charge-transfer contacts [26,27]. Although the absolute mobility values observed here are lower than those reported by Park et al. in their flux-grown samples [26], the qualitative density dependence is consistent and suggests the enhanced role of impurity scattering with decreasing carrier density.

To conclude, our data provide the first experimental confirmation of the Drude frequency response in an electrostatically gated monolayer semiconductor. Moreover, they validate the experimental methods developed for directly measuring THz conductivity in mesoscopic gated devices. Our work lays the foundation for investigating various correlated electronic states in vdW heterostructures using THz spectroscopy.

**Data availability**

The data that support the findings of this study are available from the corresponding author upon reasonable request.

**Supporting Information**

Additional details on device fabrication, measurements, and electromagnetic simulations.

**Acknowledgement**

We thank A. Y. Joe for discussions. The THz measurement was supported by the NSF award no. 2311205. The device fabrication was supported by the U.S. Department of Energy, Office of Science, Office of Basic Energy Sciences, Materials Sciences and Engineering Division under contract no. DE-AC02-05-CH11231 (van der Waals heterostructures program, KCWF16). The simulations were performed at the Molecular Graphics and Computation Facility (MGCF) at UC Berkeley and the MGCF is in part supported by NIH S10OD034382. K.W. and T.T. acknowledge support from the Japan Society for the Promotion of Science (KAKENHI grants 21H05233 and 23H02052) and the World Premier International Research Center Initiative, Ministry of Education, Culture, Sports, Science and Technology, Japan. R.Q. acknowledges support from the Kavli ENSI Graduate Student Fellowship. S.D.C. acknowledges support from the Kavli ENSI Heising-Simons Junior Fellowship.

**Author contributions**

F.W. and S.D.C. proposed the experiment; S.D.C. fabricated the devices with help from Q.F., W.Z., R.Q., Z.Z., D.A., and J.X.; S.D.C. built the experimental setup and collected the data with help from W.Z., C.U., and D.A.; S.D.C. performed the simulations and analyzed the data; T.T. and K.W. grew the hBN crystals; S.D.C. and F.W. wrote the manuscript; F.W. supervised the project.

**Competing interest**

The authors declare no competing interests.

# Supporting information

# Direct measurement of terahertz conductivity in a gated monolayer semiconductor

Su-Di Chen[1,2,3], Qixin Feng[1,3], Wenyu Zhao[1], Ruishi Qi[1], Zuocheng Zhang[1], Dishan Abeysinghe[1,3], Can Uzundal[1,3,4], Jingxu Xie[3,5], Takashi Taniguchi[6], Kenji Watanabe[7], Feng Wang[1,2,3,*]

[1] Department of Physics, University of California, Berkeley, CA 94720, USA.

[2] Kavli Energy NanoSciences Institute, University of California Berkeley and Lawrence Berkeley National Laboratory, Berkeley, CA 94720, USA.

[3] Materials Sciences Division, Lawrence Berkeley National Laboratory, Berkeley, CA 94720, USA.

[4] Department of Chemistry, University of California, Berkeley, CA 94720, USA.

[5] Graduate Group in Applied Science and Technology, University of California, Berkeley, CA 94720, USA.

[6] Research Center for Materials Nanoarchitectonics, National Institute for Materials Science, 1-1 Namiki, Tsukuba 305-0044, Japan.

[7] Research Center for Electronic and Optical Materials, National Institute for Materials Science, 1-1 Namiki, Tsukuba 305-0044, Japan.

* To whom correspondence should be addressed: fengwang76@berkeley.edu

1. Device fabrication

We start from a 500-um-thick Z-cut quartz substrate, chosen for its minimal THz absorption. Using photolithography followed by magnetron sputtering at room temperature, we first deposit a 5-nm-thick ITO gate electrode (Fig. S1, dashed rectangle) onto the substrate. After liftoff, we anneal the as-grown ITO at 250 °C for 2 minutes in a nitrogen environment to improve its stability.

Separately, we exfoliate $WSe_2$ (HQ graphene), hBN, and graphite flakes onto $SiO_2$/Si substrates. Using a Bisphenol-A polycarbonate (PC) stamp, we pick up the top hBN (43-nm thickness), monolayer $WSe_2$, few-layer graphite (Gr, used as contact to $WSe_2$), and bottom hBN (73-nm thickness) flakes sequentially at around 75 °C. The assembled stack is then released onto the ITO gate electrode at 180 °C. Afterwards, the PC residue is dissolved in chloroform at room temperature. The chip is cleaned in acetone, isopropanol, and deionized water, before dried under nitrogen flow. The hBN thicknesses are determined using AFM after completion of the device.

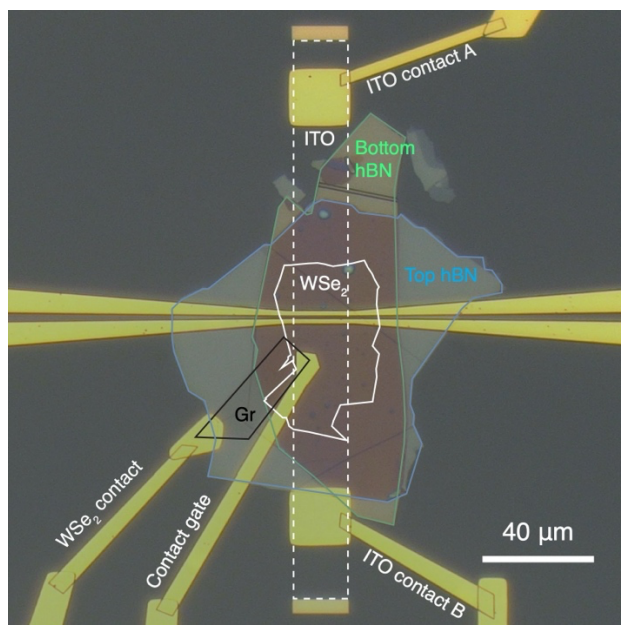

Fig. S1. Microscope photograph of central part of the device near the $WSe_2$ sample. The blue, solid white, black, green, and dashed white lines mark the boundaries of top hBN (43-nm thickness), monolayer $WSe_2$, few-layer graphite (Gr), bottom hBN (73-nm thickness), and ITO (5-nm thickness), respectively. The two-terminal resistance of ITO is measured between ITO contacts A and B, which are separated by 130 μm. The overlap between $WSe_2$ and ITO is approximately 56 μm along the vertical direction of the photograph.

With the stack in position, we use photolithography and electron-beam evaporation to define the contacts to Gr and ITO, as well as the wide sections of the CPS waveguide away from the sample and the photoconductive switches. 90 nm of Au is deposited in this step, using 5-nm-thick Cr as a sticking layer. The Au traces in the CPS are measured under an optical microscope to be 20.4-μm wide, with a 7.6-μm gap in between.

We then transfer the GaAs switches onto the reserved spots in the CPS. The switches are fabricated using the method described in Ref. [1], and the transfer is performed using a polydimethylsiloxane (PDMS) stamp cured in a Si mold. The switches are placed such that their slanted sidewalls face the propagation direction of the CPS. Another round of photolithography and electron-beam evaporation is used to connect the switches to the CPS. Before metallization, we soak the chip in 20% HCl for 3 minutes to remove the oxidized surface layer of GaAs. A 180-nm Au layer is deposited from the top, and two 60-nm layers are deposited at 45° facing the two slanted sidewalls of GaAs to ensure robust connections. A 5-nm Cr sticking layer is used in each deposition.

The final step is to fabricate the narrowed section of the waveguide. The pattern, which is designed to gradually shrink the CPS width by a factor of 10 while keeping the width-to-gap ratio constant, is defined using electron-beam lithography (EBL). During the same write, a replica of the narrowest segment is defined roughly 1 mm away from the actual CPS. After Cr/Au metallization using electron beam evaporation, this replica is inspected using SEM and the metal width (gap) is determined to be 1.98 μm (0.77 μm). This EBL/ metallization process is then repeated, with an unintentional 0.28 μm offset (measured by AFM afterwards) along the vertical direction in Fig. S1, which results in a 245-nm thick pattern with a two-step sidewall profile as shown in Fig. 2b. These measured actual (instead of designed) geometries are then used in the simulations.

In all photolithography processes, we use a double-layer resist consisting of LOR5A (Microchem) and S1818 (Shipley). The developer and liftoff solvent are MF319 (Shipley) and Remover PG (Kayaku), respectively. For the EBL processes, we use a double-layer resist formed by 495 PMMA A4 and 950 PMMA A4 (Microchem). A layer of Espacer 300Z (Snowa Denko) is coated on top to prevent charging. The electron beam energy is 100 keV. The developer is iced isopropanol-water mixture with 3:1 volume ratio, and the liftoff solvent is acetone.

2. Measurements

All measurements are performed in a vacuum chamber with optical access, where the sample mount is cooled to a base temperature of 20 K using a closed cycle cryocooler (Advanced Research Systems).

For THz measurements, the output of a femtosecond fiber laser (Menlo Systems Orange, 1040 nm, 80 MHz) is frequency-doubled in a β-BaB2O4 (BBO) crystal and used to excite the photoconductive switches. A 5X objective (Mitutoyo) is used to allow a large field of view covering both switches. All DC voltages are applied using Keithley 2614B source meters, while the AC monitoring voltage $V_m$ is applied using a Rigol DG1022Z waveform generator. The delay modulation is implemented using a retroreflector mounted on a modulated translation stage (Thorlabs NF15AP25). The detector photocurrent is collected using a 1211 current preamplifier (DL instruments), and the lock-in measurements are performed using two SR830 lock-in amplifiers (Stanford Research Systems). The Keithley source meters are also used to measure the resistance of ITO with a bias voltage not exceeding 10 mV.

For optical measurements, the same objective is used. A halogen lamp illuminates the entire sample, and a pin hole is used to filter out reflected light from a single spot in a confocal geometry. The selected beam is then collected by a home-built spectrograph with a thermoelectrically cooled camera (Pixis 100, Princeton Instruments). The objective is moved to access different spots on the sample.

3. Optical determination of $V_0$

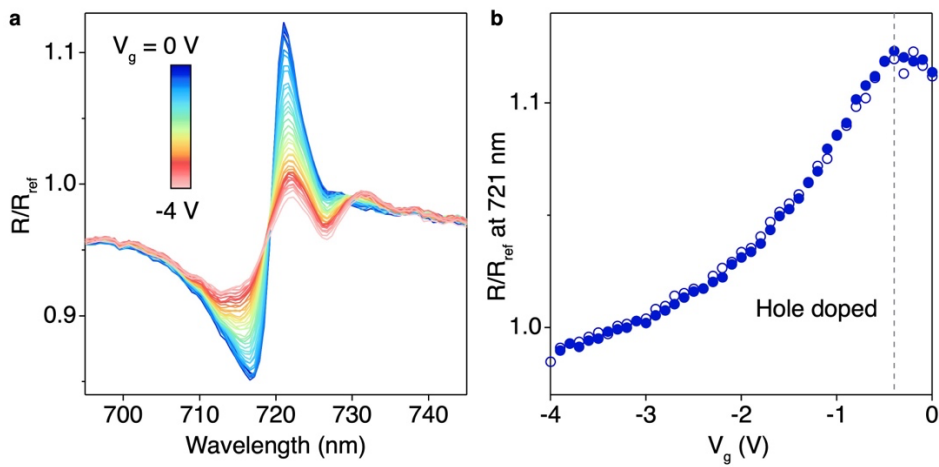

Fig. S2. Optical characterization of sample. **a**. Reflectance (R) of $WSe_2$ at different gate voltages $V_g$, normalized by that of hBN without $WSe_2$ ($R_{ref}$). Data collected in 0.1 V steps. **b**. $R/R_{ref}$ at 721 nm as a function of $V_g$. Open and closed circles represent data measured with decreasing and increasing $V_g$, respectively. As marked by the dashed vertical line, hole doping happens below -0.4 V without notable hysteresis.

For gated regions of $WSe_2$ not covered by the CPS, the onset voltage for hole doping $V_0$ can be determined using optical measurements [2]. Fig. S2 shows the evolution of the A exciton resonance in $WSe_2$ as a function of $V_g$. With decreasing $V_g$, the peak at 721 nm begins to drop at -0.4V, indicating the onset of hole doping. We have checked a few spots on the sample and the variation of $V_0$ is around 0.1 V.

4. ITO sensing of $WSe_2$ doping state

From ITO test samples sputtered on $SiO_2$/doped Si chips, we have learned that (1) the contact resistance between ITO and Cr/Au is negligible in comparison with the ITO sheet resistance, and (2) $\sigma_{ITO}$ changes linearly with gate voltage (applied to the Si backgate in these tests) and thus carrier density when the gating-induced density change $\Delta n_{ITO} \ll n_{ITO}$.

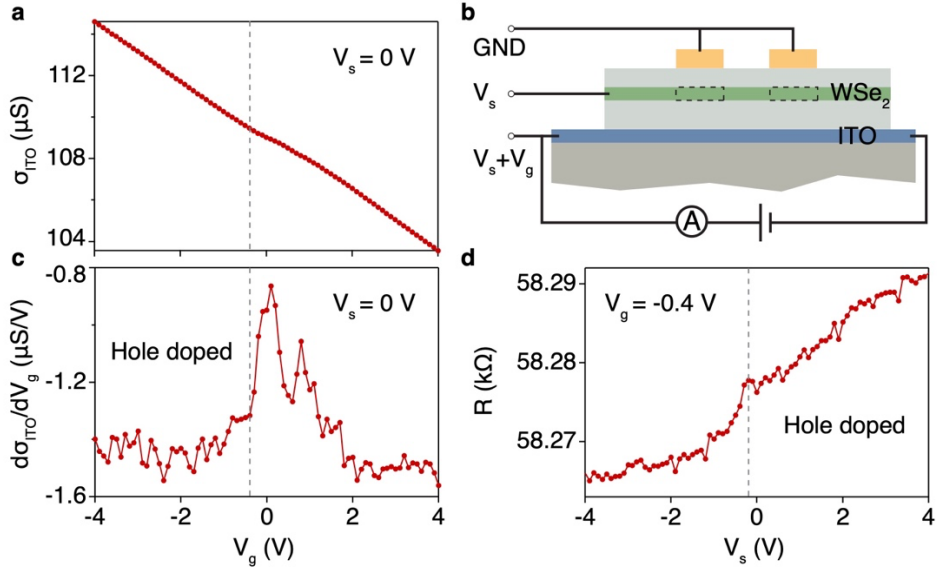

Fig. S3. ITO sensing. **a**. Sheet conductance of ITO in the region overlapping with WSe2 as a function of $V_g$. $V_s$ is fixed at 0 V during measurement. Data are converted from two-terminal resistance measurements according to the geometry illustrated in Fig. S1. As the waveguide traces are much narrower than the full overlapping region, effects from the dual-gated regions are ignored. **b**. Schematic of the resistance measurement. The dual-gated regions in WSe2 are highlighted by the dashed rectangles with exaggerated width. **c**. Derivative of the data in **a** with respect to $V_g$. The change of slope in $\sigma_{ITO}$ at -0.4 V (vertical dashed line) indicates that the chemical potential of WSe2 is swept across its valence band edge. **d**. Two-terminal resistance of ITO (R) as a function of $V_s$. $V_g$ is fixed at -0.4 V during measurement. The change of slope at -0.2 V (vertical dashed line) indicates that the chemical potential of WSe2 is swept across its valence band edge in the dual-gated regions. During measurements, we observe a slow and constant drift of R with time, likely because of oxygen loss in ITO in vacuum. Nevertheless, the drift is small, and its effect is further suppressed by averaging many back-and-forth scans.

Utilizing these effects, we can determine the onset of hole doping in WSe2 by measuring the two-terminal resistance of ITO (R) in the actual device. For large negative $V_g$ values, the WSe2 is hole doped and the sheet conductance of ITO changes linearly with $V_g$ (Fig. S3a) as expected. A mobility $\mu_{ITO} \approx 40$ cm$^2$V$^{-1}$s$^{-1}$ is estimated from the slope. With increasing gate voltage, the slope changes at -0.4 V (Fig. S3c), indicating that the chemical potential of WSe2 leaves the valence band. This onset voltage agrees with the value extracted from the optical measurements in Fig. S2.

To determine the $V_s$ that aligns the WSe2 hole density in the dual-gated region (dashed rectangles in Fig.S3b) with the rest of the sample, we fix $V_g$ at -0.4 V and measure R as a function of $V_s$ (Fig. S3d). For large positive $V_s$ values where the WSe2 is hole doped, a constant slope is observed. With decreasing $V_s$, a change of slope happens at -0.2 V. This indicates that the chemical potential of WSe2 enters its bandgap, such that more electric field acts on the ITO backgate instead. Thus, by fixing $V_s$ at -0.2 V and sweeping $V_g$, the hole densities beneath the waveguide and in the rest of the gated region are approximately aligned in the THz measurements.

5. <u>Effect of $V_g$-dependent $\sigma_{ITO}$ on measuring sample conductivity</u>

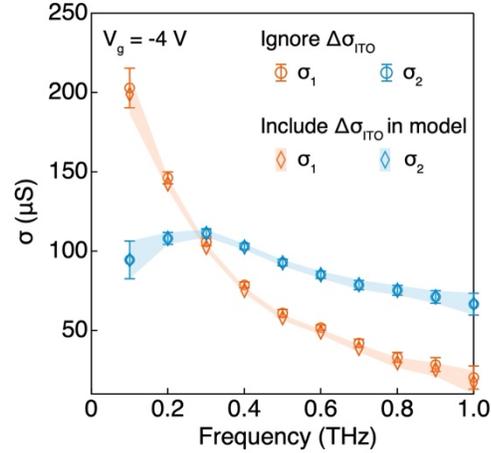

Fig. S4. Comparison of WSe$_2$ conductivity $\sigma$ obtained using two different mappings between $t/t_0$ and $\sigma$. Circles: the change of $\sigma_{ITO}$ with $V_g$ ($\Delta\sigma_{ITO}$) is ignored during the calculation of the mapping. Diamonds: $\Delta\sigma_{ITO}$ is included in calculation. The error bars and shaded regions represent 1-$\sigma$ error propagated from noise in measurements.

In the main text, we ignore the gate dependence of $\sigma_{ITO}$ because it is much smaller than the gating-induced conductivity change in WSe$_2$. This effect can also be included and compensated, by calculating $t_0$ using the measured $\sigma_{ITO}$ (Fig. S3a) at $V_g = 0$ V and t using $\sigma_{ITO}$ at the corresponding $V_g$. In Fig. S4 we compare the WSe$_2$ conductivity obtained through both treatments for $V_g = -4$ V. Including the $V_g$ dependence of $\sigma_{ITO}$ results in a slightly smaller $\sigma_1$. Although the difference is minimal here, this compensation may become important for samples with lower mobilities.

6. <u>Determine DC dielectric constant of hBN</u>

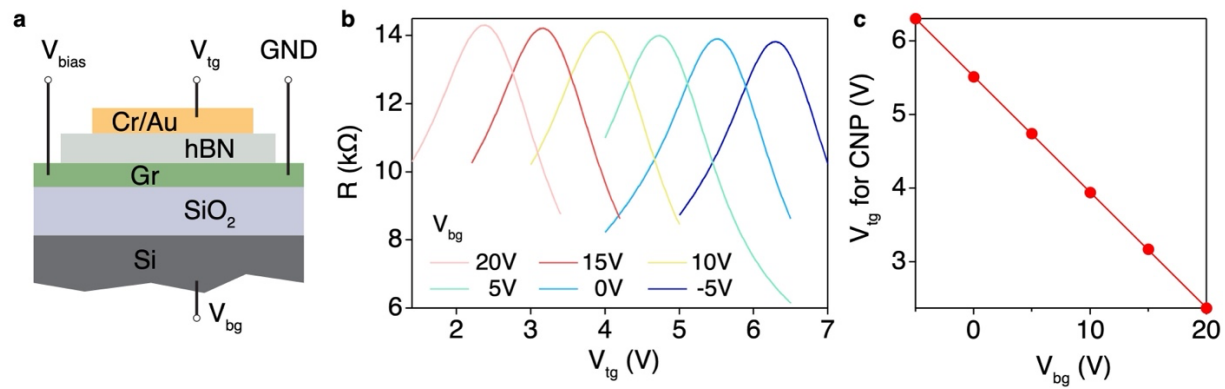

Fig. S5. Determine DC dielectric constant of hBN. **a**. Schematic of the dual-gated graphene (Gr) device. **b**. Two-terminal resistance (R) of graphene as a function of topgate voltage ($V_{tg}$) at different backgate voltage ($V_{bg}$) values. The bias voltage ($V_{bias}$) used in measurement is ±5 mV. **c**. Circles: resistance peak positions in **b** as a function of $V_{bg}$. The red line is a linear fit to the data. CNP: charge neutral point.

We use a separate dual-gated graphene device to determine the DC dielectric constant of hBN (Fig. S5). The graphene is gated through a 32-nm-thick hBN flake by a Cr/Au electrode on the top, and through 285±20 nm of thermal oxide by a doped-silicon substrate on the bottom. The hBN flake is obtained from the same batch of crystals used in the actual THz device. We measure the two-terminal resistance of graphene to determine the charge-neutral top gate voltage ($V_{tg}$) at different back gate voltages ($V_{bg}$). From the slope of $V_{tg}$ at charge neutral as a function of $V_{bg}$, we determine the DC dielectric constant of hBN $\varepsilon_r = 2.8 \pm 0.3$ using a thermal oxide dielectric constant of 3.9.

7. <u>Details of waveguide mode simulations</u>

We use the ARPACK mode solver provided in COMSOL to solve the waveguide mode in each CPS segment. The center regions we simulate have sizes larger than 400 μm × 400 μm. They are surrounded by a perfectly matched layer to allow radiative loss. A symmetry condition is enforced to reduce the computation cost. Inside the center region, the metal traces are modeled as perfect conductors. In the sample segment, $WSe_2$ is treated as a boundary, where the surface current density equals to the product of in-plane electric field and sheet conductance. The thickness of $WSe_2$ is ignored, and its width is set to 40 μm. As this width is much larger than the waveguide width, the detailed value does not affect the simulation result. The ITO is modeled in a similar way, with its width set to 50 μm. The THz dielectric constants of hBN and quartz are set to 4.67 and 4.44 for the in-plane directions, and 4.20 and 4.64 for the out-of-plane direction, respectively [3,4]. Because these dielectric constants are close to each other, changing the width of hBN does not affect the results much, even in the segments where the waveguide width is large. Thus, for segments in which hBN flakes exist, their width is set to either 80 μm (without ITO) or 50 μm (with ITO).

To ensure that the correct mode is found, we have verified that the energy carried by the mode is concentrated near the metal traces, and the current along the propagation direction is negligible inside the sample and the ITO. We have also simulated situations where analytical expressions are available [5] and verified the correctness of results.

8. <u>Effect of ITO gate electrode on the THz field</u>

In our device geometry, the ITO gate electrode inevitably reflects and screens parts of the THz field. However, when $\sigma_{ITO}$ is small, these effects are weak enough such that the sample response can still be measured with high sensitivity.

To illustrate this, we calculate the reflection (r) and transmission (t) coefficients of the THz field through the waveguide as functions of sample conductance σ, while varying $\sigma_{ITO}$ from 0 to 10 mS. As shown in Fig. S6a and b, |r| and |t| for $\sigma_{ITO}$ = 0.1 mS (squares) closely follow their counterparts at $\sigma_{ITO}$ = 0 (circles), indicating that the ITO gate electrode only weakly reflects and screens the THz field. This is no longer true with further increase in $\sigma_{ITO}$ (1 and 10 mS, triangles), where the reflection becomes substantial (Fig. S6a), the transmission drops notably (Fig. S6b), and both the absolute (Fig. S6b) and relative (Fig. S6c) changes of |t| induced by sample conductance decrease in magnitude.

In our device, $\sigma_{ITO}$ is measured to be around 0.11 mS, which only causes minimal reflection and screening of the THz field according to our simulations. We remark that these effects, although weak, are taken into account in our analysis, as we use the experimentally measured $\sigma_{ITO}$ in simulations to compute the mapping between t and σ.

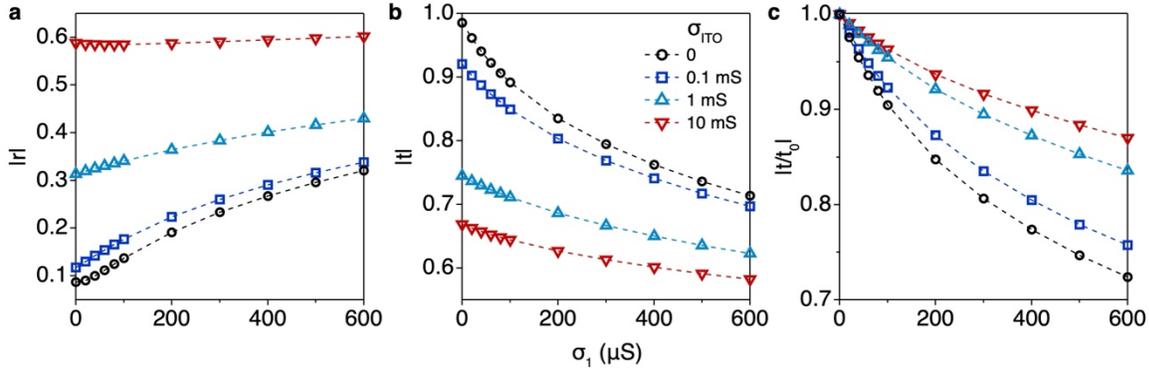

Fig. S6. Influence of $\sigma_{ITO}$ on the THz field. **a**. Magnitude of field reflection coefficient r as a function of the real part of sample conductance $\sigma_1$, calculated for $\sigma_{ITO}$ values from 0 to 10 mS as labeled in **b**. The imaginary part of sample conductance $\sigma_2$ is set to 0 and the frequency is set to 0.5 THz. **b**. Same as **a** for the transmission coefficient t. **c**. Same as **b**, with each curve normalized with respect to its value at $\sigma_1 = 0$.